# Boosting biocompatibility and mechanical property evolution in a high-entropy alloy via nanostructure engineering and phase transformations

Thanh Tam Nguyen[1,2], Payam Edalati[3], Shivam Dangwal[1,4], Karina Danielle Pereira[5]; Alessandra Cremasco[3,5], Ricardo Floriano[3,5], Augusto Ducati Luchessi[5] and Kaveh Edalati[1,2,4,*]

[1] WPI, International Institute for Carbon Neutral Energy Research (WPI-I2CNER), Kyushu University, Fukuoka 819-0395, Japan
[2] Mitsui Chemicals, Inc. - Carbon Neutral Research Center (MCI-CNRC), Kyushu University, Fukuoka 819-0395, Japan
[3] Universidade Estadual de Campinas (UNICAMP), Faculdade de Engenharia Mecânica (FEM), Brazil
[4] Department of Automotive Science, Graduate School of Integrated Frontier Sciences, Kyushu University, Fukuoka, Japan
[5] Universidade Estadual de Campinas (UNICAMP), Faculdade de Ciências Aplicadas (FCA) Brazil

High-entropy alloys (HEAs), as multi-component materials with high configurational entropy, have garnered significant attention as new biomaterials; still, their low yield stress and high elastic modulus need to be overcome for future biomedical applications. In this study, nanograin generation is used to enhance the strength and phase transformation is employed to reduce the elastic modulus of a biocompatible Ti-Zr-Hf-Nb-Ta-based HEA. The alloy is treated via the high-pressure torsion (HPT) process, leading to (i) a BCC (body-centered cubic) to ω phase transformation with $[10\bar{1}]_\omega//[01\bar{1}]_{BCC}$ and $[2\bar{1}1]_\omega//[\bar{1}2\bar{1}]_{BCC}$ through a twining mechanism, (ii) nanograin formation with a mean grain size of 20±14 nm, and (iii) dislocation generation particularly close to BCC-ω interphase boundaries. These structural and microstructural features enhance hardness, increase tensile strength up to 2130 MPa, achieve tensile elongation exceeding 13%, reduce elastic modulus down to 69 GPa and improve biocompatibility. Additionally, the HEA exhibits improved anodization, resulting in a homogenous distribution of oxide nanotubes on the surface with a smaller tube diameter and a higher tube length compared to pure titanium. These remarkable properties, which are engineered by the generation of defective nanograins and the co-existence of BCC and metastable ω phases, highlight the potential of HEAs treated using severe plastic deformation for future biomedical usage, particularly in the orthopedic sector.

*Corresponding author (Email: kaveh.edalati@kyudai.jp; Tel/Fax: +80-92-802-6744)

# 1. Introduction

Due to their excellent medical properties and ease of fabrication, metals and alloys are significantly employed in medical applications, such as implants, orthodontic appliances and fracture-fixation devices [1-3]. However, the alloys currently used in these clinical practices, like stainless steels, cobalt-chromium alloys and titanium alloys, face several challenges. These include stress shielding because of the mismatch in Young's (elastic) modulus of human bones and implants as all these materials have a high elastic modulus [4,5]. Another issue is that these materials are usually strengthened by adding toxic elements like aluminum and vanadium. The release of these elements by corrosion and friction can trigger inflammatory responses or diseases like Alzheimer's [4,5]. To overcome these problems, it is important to develop novel metallic biomaterials that do not contain toxic elements and possess high strength, low elastic modulus and good biomedical properties.

High-entropy alloys (HEAs) have recently gained considerable attention because of their exceptional properties, including high strength, ductility and biocompatibility [6-9]. Comprising at least five principal elements with a minimum entropy of $1.5R$ ($R$: gas constant), HEAs exhibit unique mechanical and functional properties through several core effects such as lattice distortion, sluggish diffusion, high mixing entropy and so-called cocktail effect [10]. These materials usually contain single- or dual-phase solid solutions, often with face-centered or body-centered cubic crystals (BCC or FCC) [11,12]. Various HEAs, such as TiZrNbTaMo [13], $Ti_{2.6}$ZrNbTaMo [14-17], $Ti_{1.7}$ZrNbTa$Mo_{0.5}$ [14-17], $Ti_{1.5}$ZrNbTa$Mo_{0.5}$ [14-17], $Ti_{1.4}Zr_{1.4}Nb_{0.6}Ta_{0.6}Mo_{0.6}$ [14-17], TiZrHfMo$Cr_{0.2}$ [18], TiZrHfMo$Cr_{0.07}Co_{0.07}$ [18], TiZrNbTaFe [19], TiZrHfNbTa [20] and $Ti_{1.5}$ZrH$f_{0.5}$NbT$a_{0.5}$ [21], have been developed for biomaterial applications. These bio-HEAs incorporate non-toxic and non-allergenic elements in their chemical composition, and they are typically synthesized through arc melting, mechanical alloying followed by sintering or additive manufacturing [14-21]. These techniques usually produce coarse-grained materials, which would be detrimental to the mechanical properties. Moreover, the mentioned alloys have high elastic modulus which makes them inappropriate for orthopedic applications.

In recent years, severe plastic deformation (SPD) processes have been employed to improve the mechanical properties and biocompatibility of metallic material [22-24]. The enhancements in properties are often due to the generation of defects and ultrafine-grained structure [22]. Among various SPD methods such as equal-channel angular pressing, high-pressure torsion (HPT), accumulative roll-bonding, multi-directional forging and twist extrusion [23], HPT exhibits the highest effectiveness for grain refinement at room temperature in various materials, including biomaterials [23]. Besides efficient grain refinement, the HPT method is capable of controlling phase transformation in various types of metallic alloys and non-metallic compounds [26,27]. These phase transformations were occasionally reported in some HPT-processed HEAs, such as carbon-doped AlTiFeCoNi [28]. Additionally, HPT was employed to mix elemental powders at the atomic scale and synthesize various materials, including biocompatible alloys, bio-composites and bio-HEAs [3,29]. Despite all these benefits, there have been few studies in using the potential of HPT to produce nanostructured HEAs with enhanced strength, low modulus of elasticity and good biocompatibility. Moreover, although anodization is usually used to grow an oxide layer (mainly in the form of nanotubes) on the surface of biomaterials containing titanium to modify their surface characteristics (such as wettability, roughness and cell-implant interaction) [30-32], the anodization behavior of biocompatible HEAs has not been investigated.

This study develops a nanostructured biomaterial using the HPT method, employing fully non-toxic and allergy-free elements titanium, zirconium, hafnium, niobium and tantalum [33]. The HPT process is used to induce various defects like twins and dislocations and achieve a phase transition to the ω crystal structure from the starting BCC crystal structure in the alloy. These structural and microstructural modifications result in a Bio-HEA that offers excellent biocompatibility, large strength, good plasticity and a reduced modulus of elasticity, and is liable to modify the surface by anodic oxide nanotubes. Among various HEAs developed for biomedical applications, TiZrNbTaHf, as the most famous HEA with the BCC structure, is selected as a model material in this study, because the alloy stands out due to its unique combination of non-toxicity, phase stability and structural transformability [34]. Moreover, unlike some other reported bio-HEAs such as TiZrNbTaMo [13–17] and TiZrNbTaFe [19], the TiZrNbTaHf alloy avoids potentially cytotoxic or corrosion-prone elements like molybdenum and iron.

## 2. Materials and Experimental Procedures
### 2.1. Synthesis and Processing

An ingot of Ti-Zr-Hf-Nb-Ta-based alloy with equal atomic fraction of elements was synthesized using the arc-melting method from a mixture of high-purity metal pieces of titanium

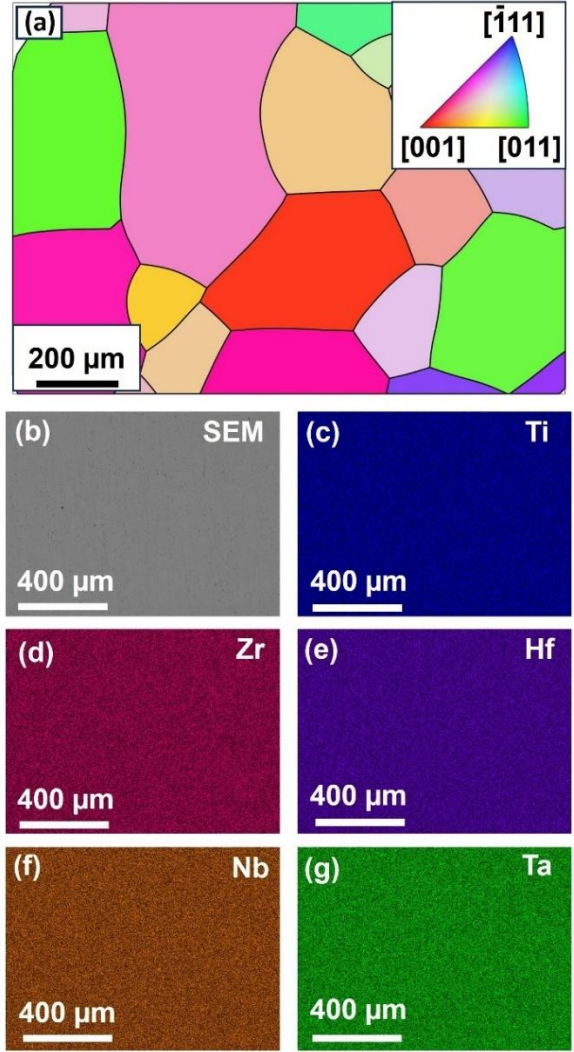

Figure 1. Coarse grains and uniform distribution of elements in Ti-Zr-Hf-Nb-Ta-based high-entropy alloy after homogenization. (a) EBSD orientation maps obtained with 10 μm step size and (b-f) EDS elemental mapping for homogenized alloy.

(99.9%), zirconium (99.7%), hafnium (99.7%), niobium (99.9%) and tantalum (99.9%) under an argon atmosphere gettered with titanium. For homogenizing the chemical distribution of the five elements, the ingot was seven times re-melted. The resulting ingot was then cut into discs having 5 mm radius and 0.8 mm thickness by employing an electron-discharge machine and polished to a smooth surface. The discs were sealed under vacuum in quartz tubes, homogenized for 60 min at 1473 K and rapid-cooled in ice water. As shown in Fig. 1, employing a scanning electron microscope (SEM) in (a) EBSD (electron back-scatter diffraction) and (b) EDS (energy dispersive

X-ray spectroscopy) modes, the homogenized disc had large grains with a 350 µm mean size and a homogeneous distribution of elements. Homogenized disc samples were treated by the HPT technique for $N = 1/8$, 1 and 10 turns at ambient temperature with an applied pressure of 6 GPa and a deformation rate of 1 rpm and characterized using different techniques. As a reference biomaterial, pure titanium with 99.9% purity was also used. Two sets of pure titanium discs were prepared by (i) annealing at 1073 K for 1 h, and (ii) HPT processing for 5 turns under 6 GPa at ambient temperature.

## 2.2. Characterization

The discs, prior and after HPT treatment, were characterized using multiple methods as illustrated in this section.

The discs were first polished mechanically to produce mirror-like surfaces. X-ray diffraction (XRD) was carried out using a Cu K$\alpha$ radiation source to investigate the crystal structure. The XRD parameters included a 40 mA anode current, a 45 kV acceleration voltage, a 0.05º step of scanning and a 2 º/min speed of scanning. A PDXL software was employed to analyze the XRD patterns.

For the nanostructural analysis, (scanning-)transmission electron microscopy (TEM and STEM) was carried out using a microscope JEOL ARM 200 F. The circular samples (1.5 mm in radius) were cut from the edge of discs for TEM. These circular samples were first polished to reduce their height to 100 µm, and later electrochemically polished employing a twin-jet electropolishing system at 463 K with 20 V using a chemical solution containing perchloric acid, butanol and methanol with volume percentages of 5, 30 and 65%, respectively. The last polishing to achieve thin foils was done by ion milling for 30 min employing argon at a voltage of 5 kV with ±5º polishing angles. The thin foils were investigated using dark-field and bright-field micrographs, selected area electron diffraction (SAED), EDS elemental maps and high-angle annular dark-field (HAADF) micrographs.

Microhardness measurements were performed on the upper surface of HPT-processed samples using a Vickers hardness testing machine. Measurements were taken at five radial directions at positions of $r = 0.1$, 1.0, 2.0, 3.0 and 4.0 mm from the disc center. The indentation force and times were 0.5 kgf and 15 s, respectively. To examine the variations of hardness against strain, the shear strain in indented points was estimated based on the relationship of $\gamma = 2\pi rN / h$ ($r$: radial distance, $N$: HPT anvil rotations, $h$: disc height).

To estimate the elastic modulus ($E$), nanoindentation tests were conducted on the mirror-like polished surfaces of discs at room temperature utilizing a nanoindenter testing machine having a Berkovich diamond indenter. The indentation was made using an applied force of 0.2 N, a loading speed of 0.4 N/min and an indentation time of 15 s. For comparison, the nanoindentation tests were also conducted for pure titanium (99.9%) under two different conditions: (i) annealed for 1 h at 1073 K and (ii) treated with 5 turns of HPT under 6 GPa at ambient temperature.

To examine mechanical properties, dog-bone-shaped samples with a cross-section of 0.6×0.6 mm$^2$ and a gauge length of 1.5 mm were cut from discs utilizing electric discharge machining. The gauge of these samples was 2 mm away from the center of the discs. The tensile test was conducted utilizing a 5.5×10$^{-4}$ s$^{-1}$ initial pulling rate.

## 2.3. Anodization Test

To evaluate the effect of HPT processing on the formation and morphology of the anodic oxide film, the anodization was conducted. Three electrolyte solutions were used for anodization, including (I) 0.3% NH$_4$F + 10% H$_2$O + ethylene glycol, (II) 0.35% NH$_4$F + 5% H$_2$O + glycerol, and (III) 0.25 M NH$_4$F + 25% H$_2$O + glycerol. The potential was set at 20 V and the anodization time was 1 h under stirring. The morphology analysis of the nanostructured layer was carried out by SEM operated at 20 kV from both surface and cross-sectional sides.

## 2.4. Biocompatibility Test

Biocompatibility encompasses various aspects such as cytocompatibility, immunogenicity, genotoxicity and long-term in vivo effects. In this study, the authorsspecifically study in vitro cytocompatibility using the MTT assay on MC3T3-E1 preosteoclast cells, which is widely applied in biomaterials research for biocompatibility. The assay provides important information on the mitochondrial activity of viable cells [9,24,29], although other biocompatibility tests are needed in the future to fully evaluate biocompatibility in a broader sense. The MTT assay was employed to examine cellular metabolic activity which is a sign of viability and proliferation of cells. In this technique, a tetrazolium salt known as MTT (3-(4,5-dimethylthiazol-2-yl)-2,5-diphenyltetrazolium bromide) with a yellow color is reduced by active cells to formazan with a purple color and the activity is examined by this change in color. For these tests, the cell line of mouse preosteoblast (MC3T3-E1) was kept in α-MEM (α minimum essential medium) accompanied with fetal bovine serum (10%), L-glutamine (2 mM), penicillin (100 U/mL) and streptomycin (100 μg/mL). These cells in α-MEM were kept at 310 K in a humidified incubator with 5% $CO_2$. Cells with 80% of confluence were collected by trypsinization and counted utilizing an automatic counter (Countess II). Later, 60 μL of suspension containing $1\times10^5$ cells were poured on the disc surfaces, previously sterilized by autoclaving and placed in a 24-well plate (one sample in each well of 1.9 $cm^2$). The samples were kept in an incubator for 2 h, followed by adding 1 mL of α-MEM to each well and returning the 24-well plate to the incubator for the time of interest. The activity of cells exposed to growth on the discs was assessed by MTT assay. After plating for 48 h, the culture medium was removed, followed by adding 0.5 mg/mL of MTT to the cells and keeping at 310 K for 3 h in a humidified incubator with 5% $CO_2$. After incubation, for dissolving MTT, the medium was discharged and 0.25 mL DMSO (dimethyl sulfoxide) was added to wells. The light absorbance was measured utilizing a scanning spectrophotometric multiwell plate reader at a wavelength 570 nm. Statistical analyses were carried out utilizing One-Way ANOVA and Student's t-test. To achieve the standard deviation of measurements, the measurements were performed in three independent tests with cells on separate passages. A SigmaStat software was used for statistical analyses and changes were statistically regarded as meaningful once probabilities (*p*) were less than 0.05.

## 3. Results
### 3.1. Structure and Microstructure

Fig. 2 displays the diffraction patterns of homogenized material and samples treated using HPT for 1/8, 1 and 10 rotations at two different magnifications. The XRD pattern of homogenized TiZrHfNbTa alloy reveals the peaks indexed as (110), (200), (211), (220) and (310), which are characteristic of BCC crystal structure, confirming the presence of a single phase of BCC (also known as the β phase [33]). The lattice parameters of BCC are measured as 0.3409 nm, consistent with previous findings [34,35]. The XRD peaks show some asymmetry, a fact that is observed in many arc-melted HEAs due to compositional gradients [13,14]. Alloys processed by HPT also exhibit the BCC phase, but a small peak appears at 38.7º, particularly after 1/8 HPT turns, which should correspond to the (002) plane of the ω phase. Moreover, XRD peaks become broader, indicating a decrease in crystallite size, the generation of dislocations and an increase in lattice strain, induced by HPT [36]. Examination of XRD profiles by Rietveld analysis suggests that while the lattice parameter of 0.3409 nm remains reasonably unchanged after HPT processing, crystallite size decreases to 34 nm after $N = 1/8$, 21 nm after $N = 1$ and 20 nm after $N = 10$, and lattice strain increases to 0.31% after $N = 1/8$, 0.45% after $N = 1$ and 0.54% after $N = 10$. It should be noted that the Rietveld analysis suggests a large crystallite size and a small lattice strain for the homogenized sample, but these values are not quantitatively reliable because of the invalidity of the kinematic theory of scattering for coarse grains.

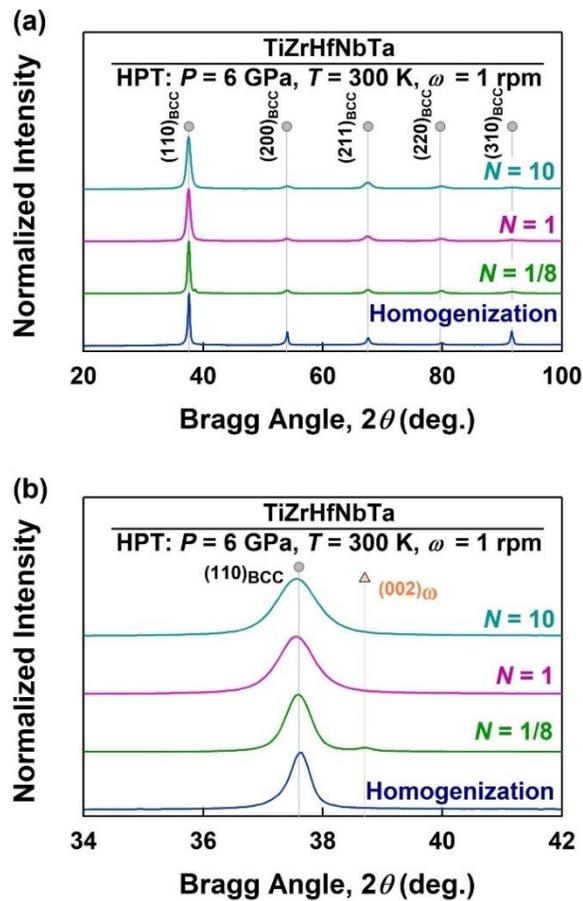

Figure 2. Lattice strain and XRD peak broadening after severe plastic deformation in Ti-Zr-Hf-Nb-Ta-based high-entropy alloy. XRD patterns of homogenized material and samples treated using HPT for $N = 1/8$, 1 and 10 turns at two different magnifications.

Fig. 3 represents TEM bright- and dark-field images and SAED profiles for samples after homogenization and after 1/8, 1 and 10 HPT turns. Examination of Fig. 3 indicates five important issues. (i) The homogenized sample shows the presence of only one grain in the selected area in Fig. 3a and 3c with a dotted SAED pattern as shown in Fig. 3b in good agreement with EBSD analysis. (ii) After HPT processing for $N = 1/8$ (Fig. 3d-g), nanotwin-like features are introduced to the alloy. The generation of nanotwins by SPD treating or post-SPD annealing was reported in various materials like copper [37,38], nickel [39], stainless steel 316 L [40], Fe-25Mn alloy [41], TWIP (twinning-induced plasticity) steels [42] and HEAs [23]. (iii) The ω phase is detected in the Ti-Zr-Hf-Nb-Ta-based alloy after HPT processing as shown in the SAED image in Fig. 3e. Detailed analysis confirms that the ω phase is initially formed in twin-like features such as the one observed in Fig. 3f. The ω-phase formation by HPT treatment was previously documented in pure titanium [43,44], pure zirconium [45] and Ti-V-Cr alloys [46] using TEM analysis. Phase transformation from BCC to the ω phase after deformation at cryogenic temperatures was also previously observed in Ti-Zr-Hf-Nb-Ta-based alloy by TEM [47]. Phase transformations to metastable phases were suggested to contribute to the enhancement of ductility in HEAs [48]. (iv) SAED analyses present distinct changes from a dot pattern to ring patterns at large HPT rotations such as $N = 1$ (Fig. 3i) and $N = 10$ (Fig. 3l), indicating the formation of numerous nanograins with random orientations. Examination of white areas of dark-field micrographs of Fig. 3j and 3m indicates the formation of nanograins after HPT processing. The grain sizes decrease from $350\pm120$ μm for the homogenized sample (measured by EBSD for about 15 grains), to $21\pm15$ nm after $N = 1$ HPT turn and further refine to a mean size of $20\pm14$ nm by HPT processing for $N = 10$ turns. It should be noted that the grain sizes for HPT-processed samples were determined from TEM dark-field images by measuring the orthogonal dimensions of white regions for over 50 grains. It should be noted that the grain sizes for $N = 1/8$ could not be measured because they were larger than the view area of the TEM used. The grain size levels after HPT processing for $N = 1$ and 10 are comparable to grain sizes observed in a few dual-phase HEAs [49] or ceramics [23], but it is considerably lower than those achievable in HPT-treated metals (usually in the submicrometer range) [23,50].

High-resolution lattice images in Fig. 4 and Fig. 5 further illustrate the microstructural features of Ti-Zr-Hf-Nb-Ta-based alloy after processing at low strains ($N = 1/8$) and high strains ($N = 10$), respectively. After HPT processing for $N = 1/8$, despite relatively large grains, numerous dislocations are detected. As illustrated in Fig. 4, the space between such dislocations is quite close, suggesting a high dislocation density in agreement with the broadening of XRD peaks shown in Fig. 1. As illustrated in Fig. 5a, the ω phase still exists after $N = 10$ turns, but it is now in the form of nanograins and not twin-like features. Examination of lattice images after the HPT treatment for 10 rotations also confirms the co-existence of the BCC phase (Fig. 5b and 5c) and the ω phase (Fig. 5d and 5e). The co-presence of these two phases generates a large fraction of interphase boundaries. Some of these interphase boundaries are almost coherent, as shown in Fig. 5d, with an orientation relationship of $[10\bar{1}]_\omega//[01\bar{1}]_{BCC}$ and $[2\bar{1}1]_\omega//[\bar{1}2\bar{1}]_{BCC}$, suggested by fast Fourier transform (FFT) diffractogram of Fig. 5e. Some other interphase boundaries, like the one shown in Fig. 5f do not show a distinct coherency. In addition to interphase boundaries, dislocations are still visible, particularly close to interphase boundaries, as shown in Fig. 5g and 5h. Because grain boundaries usually act as a sink for dislocations in nanograined metals [51], the presence of dislocations is rather unusual in this severely deformed HEA. The reason for this behavior should be the effect of a large fraction of solute atoms on pinning the dislocation motion in HEAs [52]. The combination of nanograins, dislocations and the ω phase should contribute to the strengthening of the alloy. Both Fig. 4 and 5 confirm the reduction of crystal size and increase in dislocation density, a fact that is in line with XRD peak broadening and Rietveld analysis.

Although experimental and theoretical thermodynamic parameters for the ω phase are not currently available in the literature, the phase is known to be thermodynamically formed under high pressures (several gigapascals, such as in HPT processing) and remains metastable after releasing pressure [43-47]. Some other parameters, such as configurational entropy of mixing, atomic size mismatch and valence electron concentration, can affect the survival of the ω phase in HEAs [53,54].

$$\Delta S_{mix} = -R \sum_{i=1}^{n} c_i \ln c_i = -Rn(\frac{1}{n} \ln \frac{1}{n}) \quad (1)$$

$$\delta = \sqrt{\sum c_i (1 - \frac{r_i}{\bar{r}})^2} \times 100 \quad (2)$$

$$VEC = \sum c_i VEC_i \quad (3)$$

where $\Delta S_{mix}$ is the configurational entropy of mixing, $\boldsymbol{\delta}$ is atomic size mismatch, VEC is valence electron concentration, $R$ is the gas constant, $\boldsymbol{c_i}$ is the atomic fraction of elements, $n$ is the number of elements, $r_i$ is the atomic radius (Ti: 147 pm, Zr: 160 pm, Hf: 159 pm, Nb: 146 pm, and Ta: 146 pm), $\bar{r}$ is the average atomic radius. The calculated configurational entropy of mixing, atomic size mismatch and valence electron concentration are $1.61R$, 6.6% and 4.4, respectively. The high configurational entropy of mixing favors the formation of solid solution phases and suppresses chemical changes by phase transformation [10-12]. The large atomic size mismatch promotes lattice distortion, which may facilitate the presence of distorted metastable phases such as ω after HPT processing. Furthermore, the valence electron concentration (VEC ≈ 4.4) falls within the range that typically stabilizes BCC structures in refractory HEAs, and this should be the main reason that the BCC phase remains the main phase even after HPT processing [53,54].

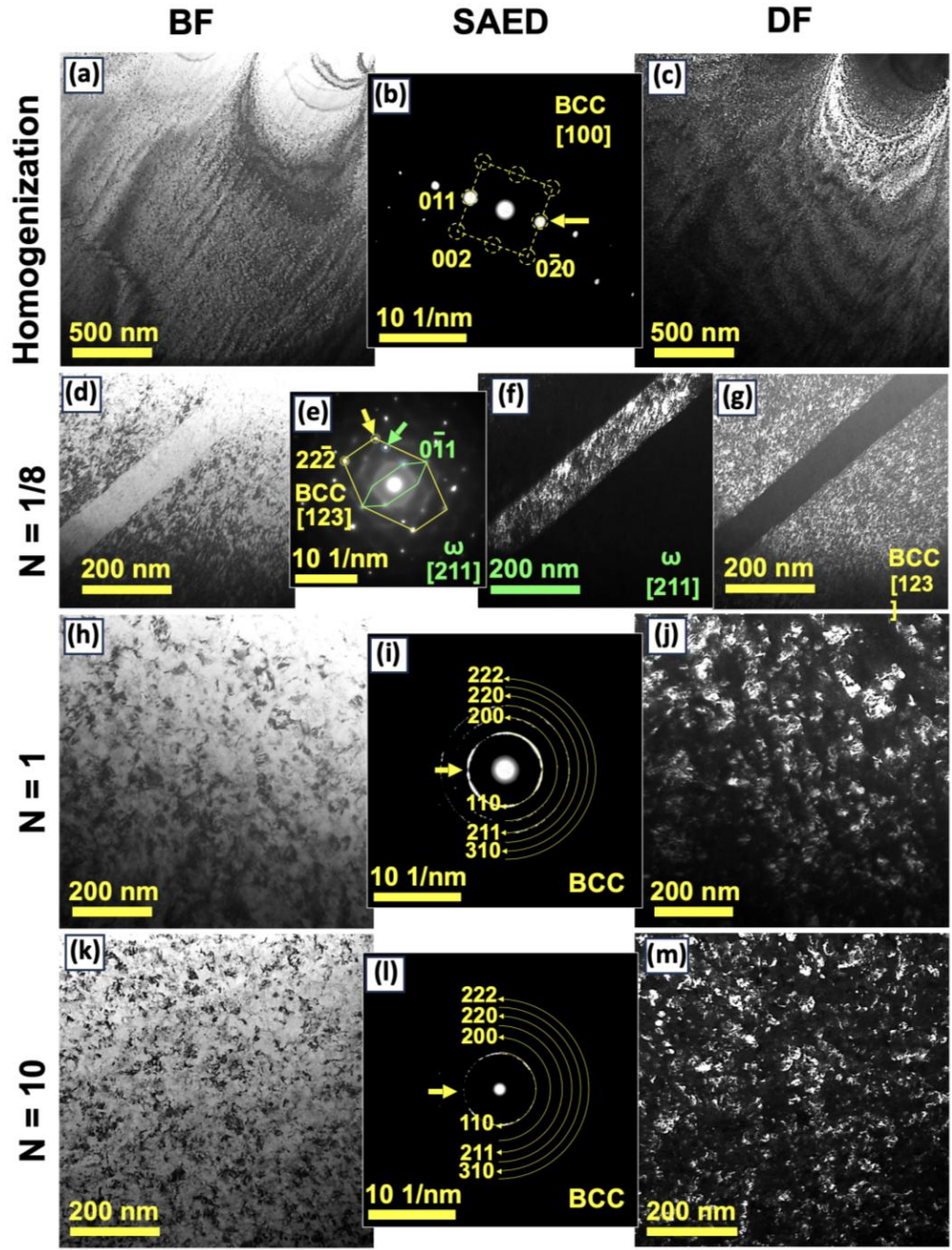

Figure 3. Strain-induced refinement of grains and BCC to ω phase transformation in Ti-Zr-Hf-Nb-Ta-based high-entropy alloy. TEM microstructural examination of samples after (a-c) homogenization and following HPT treatment with (d-g) $N = 1/8$, (h-j) $N = 1$ and (k-m) $N = 10$ rotations using (a, d, h, k) bright-field (BF) images, (b, e, i, l) SEAED analysis and (c, f, g, j, m) dark-field (DF) images. Note that dark-field micrographs were achieved utilizing diffraction beams pointed in SAED profiles by arrows.

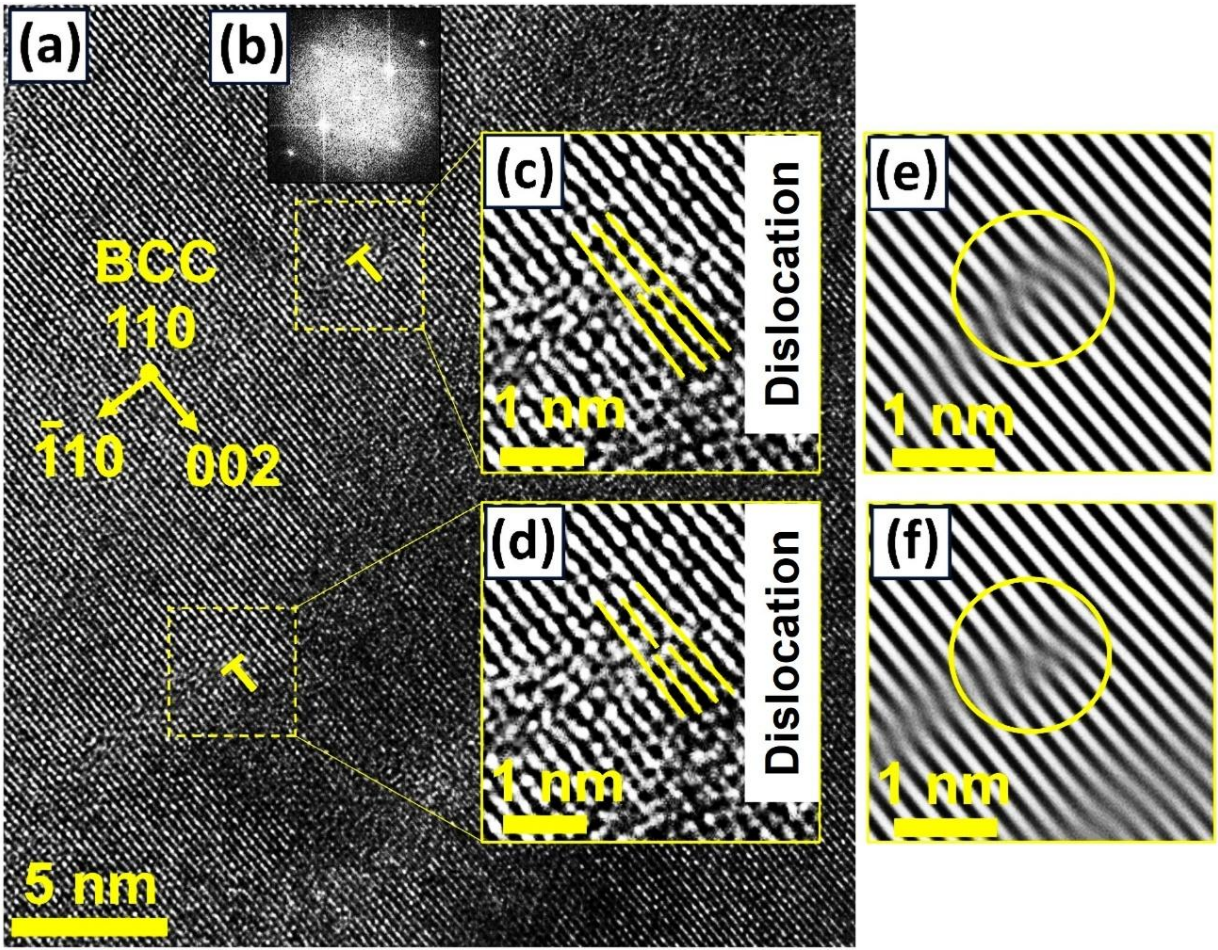

Figure 4. Strain-induced dislocation generation in Ti-Zr-Hf-Nb-Ta-based high-entropy alloy. (a) TEM high-resolution micrograph, (b) relevant FFT diffractogram, (c, d) lattice images of two dislocation-type defects and (e, f) reconstructed images of two dislocations using inverse FFT from (110) spot for disc proceed using HPT with 1/8 rotations. T marks in (a) indicate dislocations.

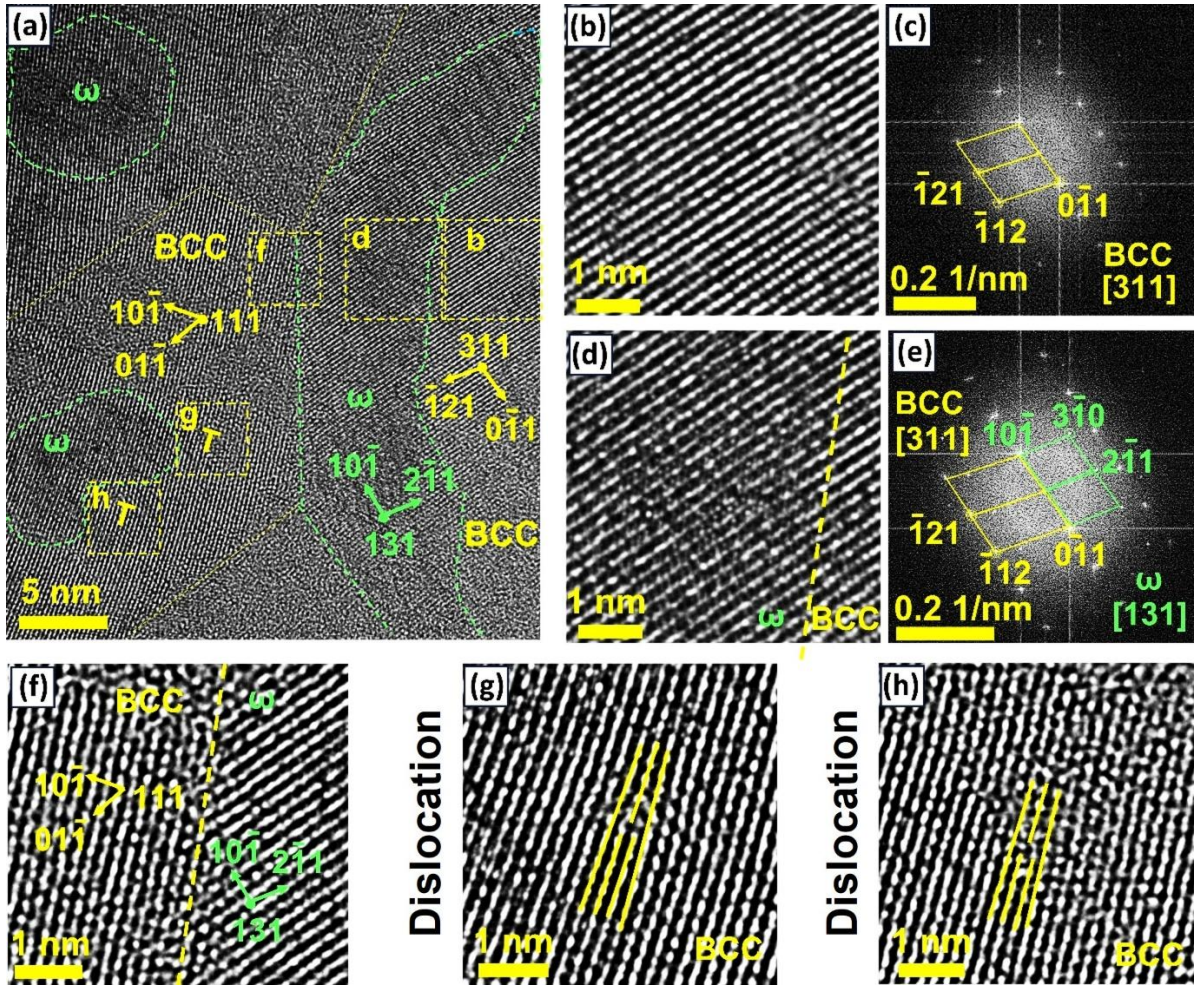

Figure 5. Strain-induced formation of BCC-ω interphases in nanograined Ti-Zr-Hf-Nb-Ta-based high-entropy alloy. (a) TEM high-resolution micrograph, (b, c) lattice image of BCC and relevant FFT, (d, e) lattice image of coherent BCC-ω interphase, (f) lattice image of incoherent BCC-ω interphase, (g, h) lattice images of two dislocations formed close to BCC-ω interphases for the disc treated by HPT with $N = 10$. Lattice images were taken from the regions shown by squares in high-resolution micrograph of (a). T marks in (a) indicate dislocations.

## 3.2. Mechanical Properties

Hardness measurements, often used for rapid microstructural evaluation during HPT processing [51,55], are plotted in Fig. 6 versus (a) distance from the center of discs and (b) shear strain. In the homogenized HEA, the hardness is uniformly distributed along the disc radius and the average hardness is 390±10 Hv. After HPT processing, the hardness rises with increasing radial distance from the center of discs as well as with increasing the HPT rotations (Fig. 6a), reaching a maximum of 500 Hv in the alloy processed for $N = 10$. The variation of hardness versus distance from the center of discs and rotation numbers can be described more clearly by plotting the hardness data versus shear strain as illustrated in Fig. 6b. During the primary deformation stages ($N = 1/8$ and $N = 1$), hardness increases with shear strain; however, at larger shear strains ($N = 10$), a steady-state hardness is achieved, where the variations of hardness become independent of applied strain. This steady-state hardness behavior is typical in severely deformed metals and is governed by balancing the softening processes with hardening processes through grain boundary migration, dynamic recrystallization or dynamic recovery [23,50].

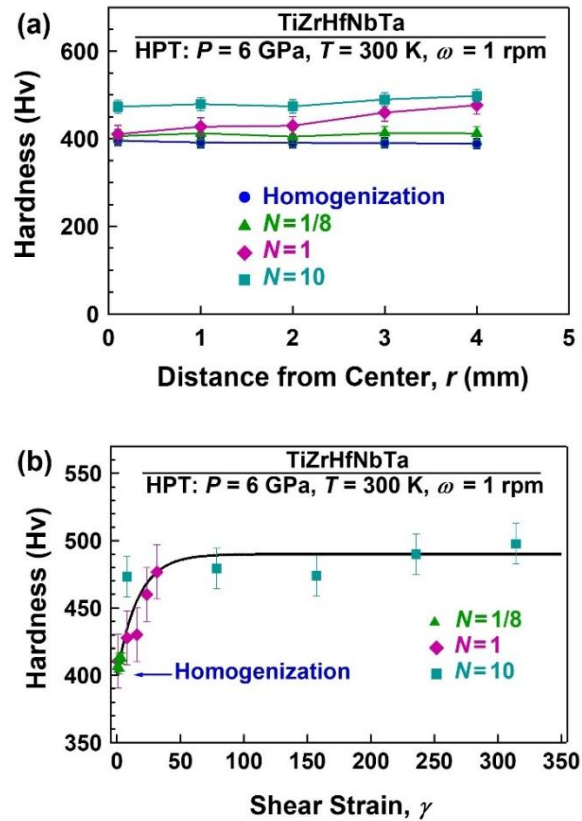

Figure 6. High microhardness after severe plastic deformation in Ti-Zr-Hf-Nb-Ta-based high-entropy alloy. Changes in Vickers microhardness against (a) distance from center of processed discs and (b) shear strain for homogenized material and discs processed by HPT with 1/8, 1 and 10 turns.

The elastic modulus determined by nanoindentation is illustrated in Fig. 7 for the homogenized material and the sample treated using HPT with $N$ = 10 rotations. For comparison, two reference samples of pure titanium, one after thermal annealing for 1 h at 1073 K and another one following the HPT process for 5 turns at ambient temperature, are also presented in Fig. 7. The homogenized HEA exhibits a modulus of elasticity of 97±4 GPa which is somehow lower than the elastic modulus of the annealed titanium sample (121±4 GPa). These measured values of elastic modulus for both HEA and titanium are consistent with the literature: 87 GPa for TiZrHfNbTa [35], 100-110 GPa for titanium [56] or 120 GPa for titanium [57]. After HPT processing, the elastic modulus for both materials decreases and reaches 69±2 GPa for the HEA and 98±2 GPa for titanium. The decrease in the elastic modulus by HPT processing cannot be explained by microstructural refinement, but it should be due to a phase transition to the ω crystal structure. Although it is generally believed that the ω phase should have a larger elastic modulus compared to the BCC and HCP phases due to its compact structure, first-principles calculations suggested that the modulus elasticity of this crystal structure can be tuned to the lower levels compared to the BCC and HCP phase in some titanium alloys [58]. The low elastic modulus of HPT-processed HEA is promising for orthopedic applications where the stress shielding effect is a concern [2].

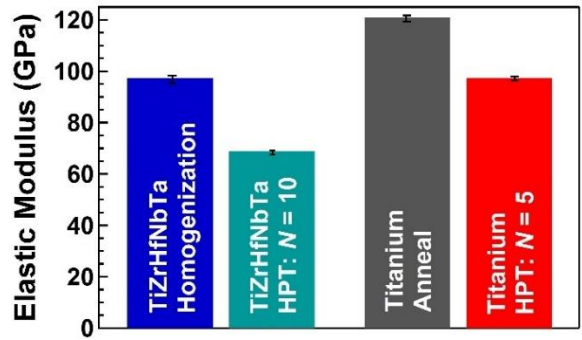

Figure 7. Low elastic modulus after severe plastic deformation in Ti-Zr-Hf-Nb-Ta-based high-entropy alloy. Examination of elastic modulus of homogenized and HPT-processed ($N$ = 10) alloy in comparison with reference annealed and HPT-processed ($N$ = 5) pure titanium.

Tensile stress-stress curves are illustrated in Fig. 8 for the homogenized material and discs treated using HPT with 1/8, 1 and 10 rotations. The homogenized sample shows brittle behavior and breaks in the elastic region under a tensile strength of 640 MPa. After processing by HPT with $N = 1/8$, both strength and elongation are considerably improved, reaching 1300 MPa and 85%, respectively. With further increasing the HPT turns to $N = 1$ and 10, the tensile strength increases to 2080 and 2130 MPa, respectively; however, elongation decreases to 13-15%. The tensile strength over 2000 MPa, which is 2-3 times larger than the tensile strength of HPT-processed titanium [44], is regarded as a high value even for severely deformed materials [23].

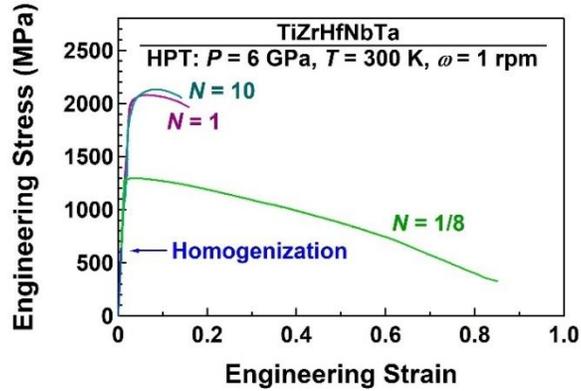

Figure 8. Ultrahigh tensile strength with good plasticity after severe plastic deformation in Ti-Zr-Hf-Nb-Ta-based high-entropy alloy. Nominal tensile stress versus strain for homogenized material and discs treated using HPT with 1/8, 1 and 10 turns.

### 3.3. Anodization

For anodization, three electrolyte solutions were used, including (I) 0.3% $NH_4F$ + 10% $H_2O$ + ethylene glycol, (II) 0.35% $NH_4F$ + 5% $H_2O$ + glycerol, and (III) 0.25 M $NH_4F$ + 25% $H_2O$ + glycerol, and anodization was conducted under a potential of 20 V for 1 h. In addition to the homogenized material and the sample after HPT treatment for 10 turns, two titanium annealed and HPT-processed samples were also anodized for comparison. In all cases, the formation of oxide nanotubes was observed, as shown as an example in Fig. 9 for the HEA discs after homogenization and HPT treatment using electrolyte I. The nanotube layer displays an extremely refined nanostructured appearance with uniform length and diameter of tubes, as illustrated in Fig. 10 from both top and cross-sectional views using SEM. As shown quantitatively in Fig. 11, the length and diameter of these nanotubes significantly depend on the type of electrolytes and the composition of the material (pure titanium or TiZrHfNbTa) rather than the processing condition (anneal, homogenization or HPT). However, it can be seen in Fig. 11 that the diameter of the tubes is smaller and their length is higher for the HEA compared to pure titanium, which is an advantage for biomedical applications [30-32]. The nanotube layer is sometimes covered with a thin oxide layer, which is dependent on the electrolyte chemical composition and material, such as in TiZrHfNbTa after anodization using electrolyte I. When electrolytes I and III are used, uniform nanostructured anodized layers are observed, but electrolyte II leads to a porous structure.

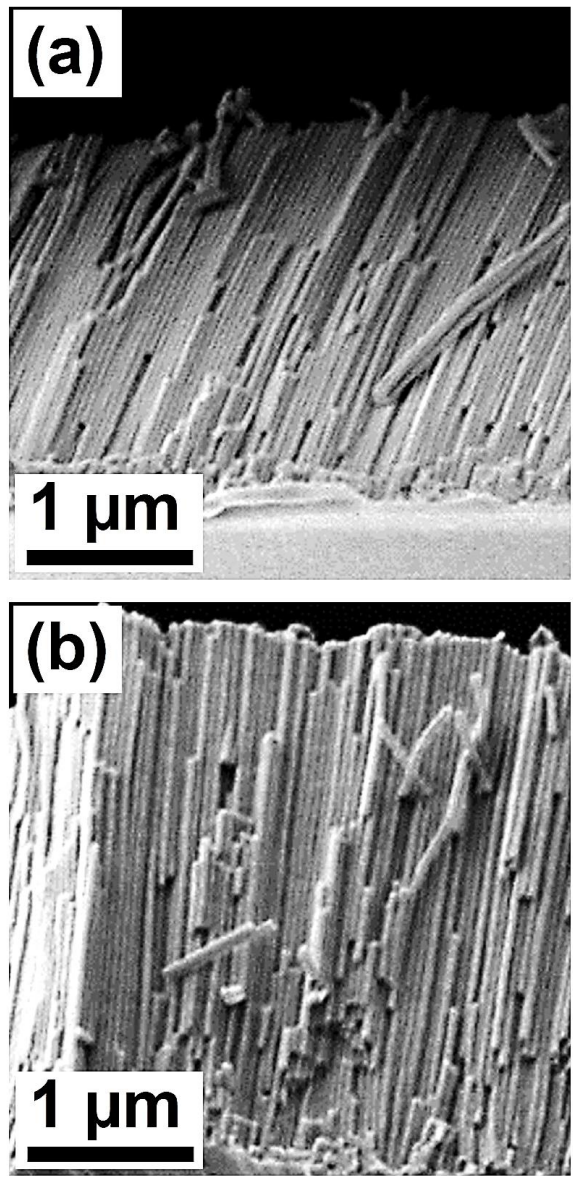

Figure 9. Formation of nanotubes on Ti-Zr-Hf-Nb-Ta-based high-entropy alloy after anodization. SEM images from cross-sectional view for homogenized and HPT-processed ($N = 10$) alloy after anodization using electrolytes (I) 0.3% $NH_4F$ + 10% $H_2O$ + ethylene glycol.

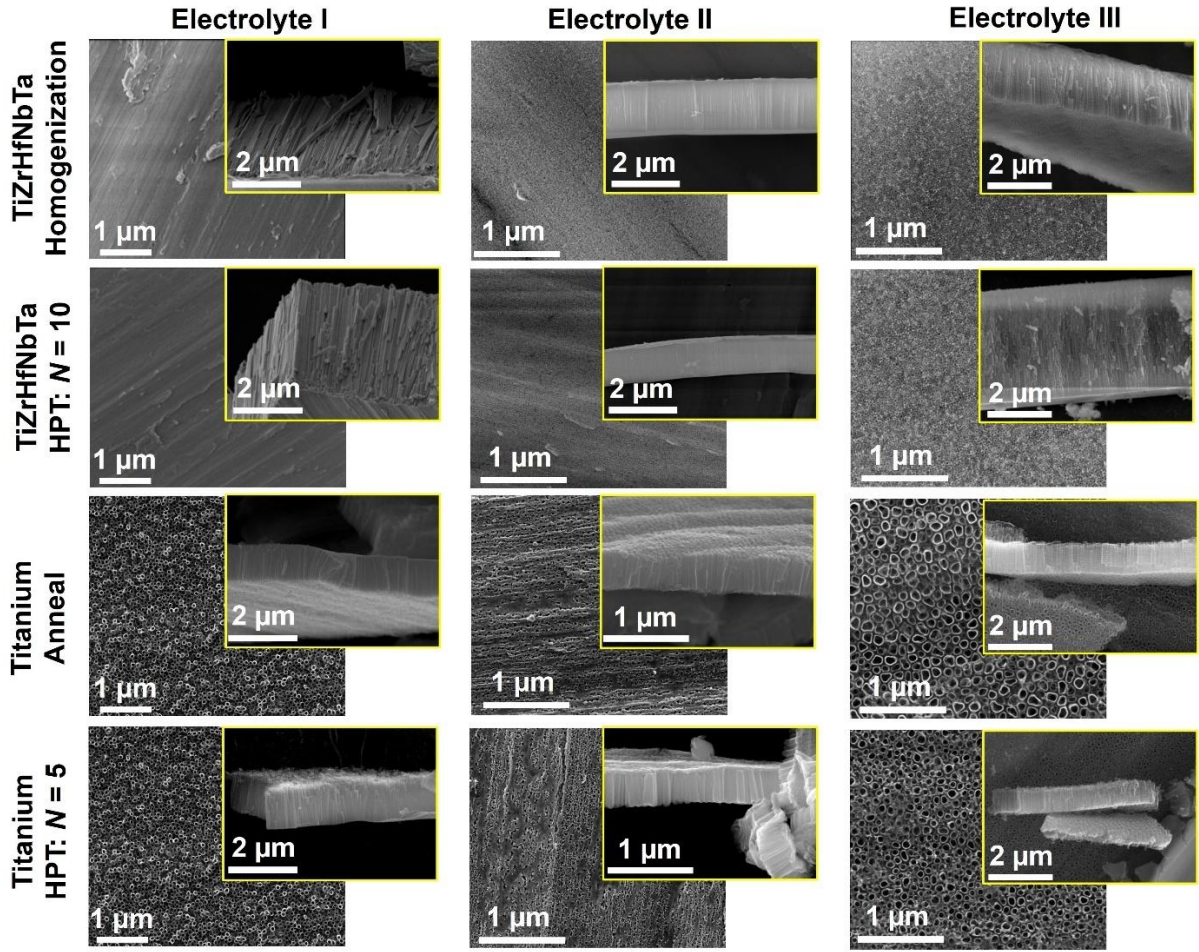

Figure 10. Formation of uniform anodized layer on Ti-Zr-Hf-Nb-Ta-based high-entropy alloy. SEM images from top and cross-section views of (a-c) homogenized and (d-f) HPT-processed ($N = 10$) alloy in comparison with reference (g-i) annealed and (j-l) HPT-processed ($N = 5$) pure titanium after anodization using electrolytes (I) 0.3% $NH_4F$ + 10% $H_2O$ + ethylene glycol, (II) 0.35% $NH_4F$ + 5% $H_2O$ + glycerol and (III) 0.25 M $NH_4F$ + 25% $H_2O$ + glycerol.

### 3.4. Biocompatibility

The biocompatibility assessment data via the MTT assay are illustrated in Fig. 12 for the homogenized material and discs treated using HPT with 1/8, 1 and 10 turns. For comparison, reference titanium discs after annealing and HPT treatment for $N = 5$ rotations are included. A larger light absorbance in Fig. 12 indicates better cell viability and higher biocompatibility. The homogenized alloy without HPT treatment exhibits lower biocompatibility compared to pure titanium. The HPT treatment results in an increase in the biocompatibility of both pure titanium and the HEA; however, the biocompatibility of HPT-treated HEA becomes better compared with the biocompatibility of reference titanium. The biocompatibility of HPT-processed HEA remains reasonably constant with increasing HPT rotations from 1/8 to 10. The improvement of biocompatibility by the SPD process is consistent with earlier reports about titanium and some biocompatible alloys [3,9,22]. These biocompatibility tests, together with increased microhardness and reduced elastic modulus of HPT-processed HEA, introduce the alloy as a potential biomaterial for orthopedic applications [1,2]. Although these biocompatibility results are promising, they are limited to early-stage in vitro viability. Further analyses, such as live/dead staining, alkaline phosphatase activity (ALP) and in vivo assessments are required to fully understand the biological performance of the material after HPT processing. Furthermore, additional assays such as bromodeoxyuridine or EdU (5-Ethynyl-2'-eoxyuridine) incorporation, Ki-67 immunostaining, or CDSE (Carboxyfluorescein succinimidyl ester) dilution should be conducted to fully evaluate cell proliferation [59,60].

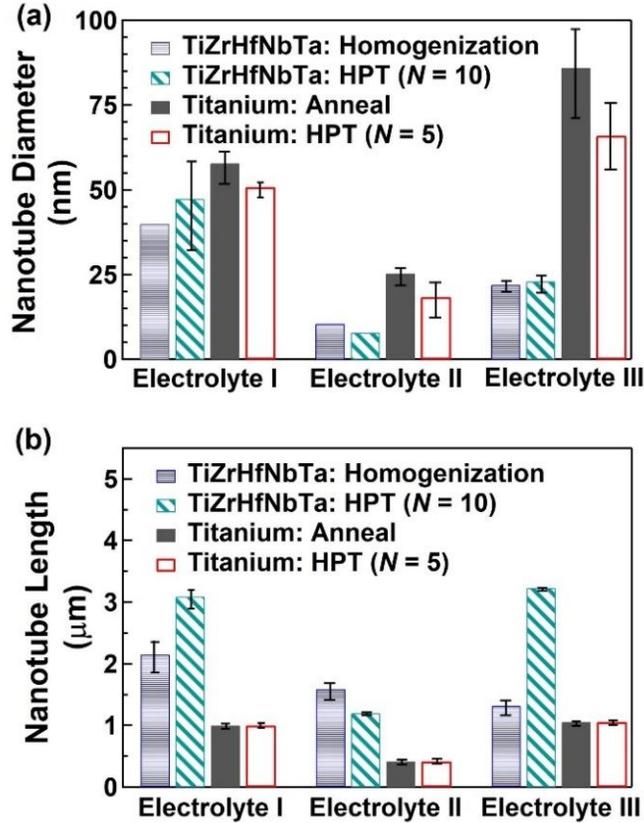

Figure 11. Smaller diameter and longer nanotube length on Ti-Zr-Hf-Nb-Ta-based high-entropy alloy compared to pure titanium after anodization. (a) Nanotube diameter and (b) nanotube length for homogenized and HPT-processed ($N = 10$) alloy in comparison with reference annealed and HPT-processed ($N = 5$) pure titanium after anodization using electrolytes (I) 0.3% $NH_4F$ + 10% $H_2O$ + ethylene glycol, (II) 0.35% $NH_4F$ + 5% $H_2O$ + glycerol and (III) 0.25 M $NH_4F$ + 25% $H_2O$ + glycerol.

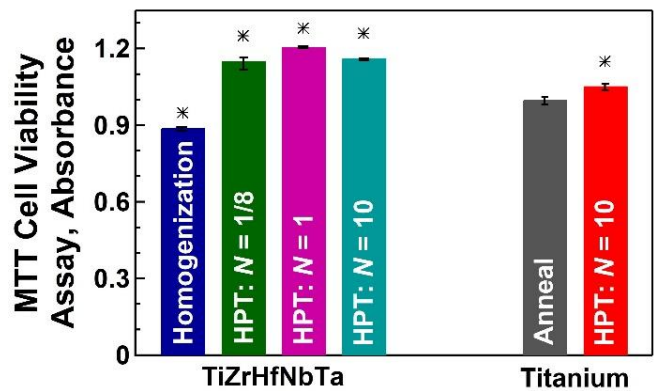

Figure 12. High biocompatibility after severe plastic deformation in Ti-Zr-Hf-Nb-Ta-based high-entropy alloy. Light absorbance at 570 nm after MTT assay for homogenized and HPT-processed ($N$ = 1/8, 1 and 10) alloy in comparison with reference annealed and HPT-processed ($N$ = 5) pure titanium.

## 4. Discussion

This research confirms the impact of the SPD by using the HPT method on the mechanical property evolution, anodization and biomedical properties of the TiZrHfNbTa alloy, a famous BCC-based HEA [34,35]. The results confirm the formation of a biocompatible HEA with superior hardness and strength, good plasticity, low elastic modulus and better biocompatibility than titanium. Two issues need further clarification here: (i) the properties of the present HEA compared with previously reported biomaterials and (ii) the major reasons for the improvement in properties of this HEA.

Regarding the first issue, Fig. 13 compares the data of microhardness and elastic modulus of the current HEA with other Ti-based biomaterials [61-66]. The HEA processed in this study shows one of the highest microhardness values of 500 Hv, whereas most other biomaterials exhibit hardness levels ranging from 250 to 350 Hv. Moreover, the elastic modulus of this material (69 GPa) is relatively low compared to many Ti-based alloys. However, this elastic modulus is still higher than the modulus of elasticity of human bones, which is within 10 to 30 GPa [67,68]. This promising integration of high hardness and reduced elastic modulus together with high tensile strength (1300-2130 MPa exceeding those reported for this HEA [6-8]) and good plasticity (13-85%) show the high potential of this material as a good biomaterial, particularly for implant applications [1-5].

Regarding the second issue, the mechanical and biocompatibility improvements in the HPT-processed TiZrNbTaHf alloy become more evident when directly compared with the homogenized HEA and reference pure titanium samples. While the homogenized HEA exhibits moderate strength (~640 MPa) and relatively low biocompatibility, the HPT-processed samples exhibit ultrahigh tensile strength levels of up to 2130 MPa with appreciable ductility (over 13%), due to the combined effects of nanograin formation, BCC to ω phase transformation, and dislocation strengthening. In contrast, pure titanium, even after HPT processing, reaches a strength below 1000 MPa. Furthermore, while the biocompatibility of homogenized HEA is initially inferior to titanium, HPT processing enhances cell viability beyond that of both annealed and HPT-processed titanium, likely due to increased surface energy, nanotopography, and lattice defects, which are known to promote cell adhesion and proliferation. Some factors affecting these features are discussed below.

The unique properties of HEAs often result from core effects of high entropy of mixing, severe distortion of lattice caused by the existence of atoms with varying atomic sizes, sluggish diffusion of atoms due to lattice potential energy fluctuations and the so-called cocktail influence, wherein each atom contributes distinct properties to the alloy [10-12]. These core effects were used earlier to develop biocompatible HEAs [13-19], including TiZrHfNbTa [20,21], for possible application as implants. However, enhancing strength and decreasing elastic modulus is a major task in using these materials for biomedical devices [1-5]. In the present work, the concept of SPD [23,25] was used to address these two issues. SPD processing via the HPT method contributes to hardening by the generation of nanograins and dislocations, a fact that was used in the past to produce various high-strength materials [23,25,50,55], including titanium [22,44] and HEAs [9,49]. The enhancement of strength in current HEA after HPT processing can be attributed to several strengthening mechanisms, although precise quantification of each mechanism is currently challenging due to a lack of sufficient data and relevant analyses for HEAs.

- First, solution hardening, which is usually significant for HEAs and results in 640 MPa strength for the homogenized sample which is larger than the strength of coarse-grained titanium.
- Second, grain boundary strengthening through the Hall-Petch mechanism, which should be significant due to the formation of nanograins with an average size of $d = 20$ nm, as observed via TEM. If the Hall-Petch strengthening ($\sigma_{HP}$) is evaluated through the equation $\sigma_{HP}$ in MPa = $942 + 270$ ($d$ in μm)$^{-1/2}$, suggested in Ref. [69], it reaches $\sigma_{HP} = 2850$ MPa, which is higher than the strength of the current HPT-processed HEA. This discrepancy should be due to the inaccuracy of the suggested Hall-Petch parameters and/or the contribution of the inverse Hall-Petch mechanism at the nanometer level [70].
- Third, the presence of high dislocation density, particularly at BCC-ω interphase boundaries, contributes to dislocation strengthening.
- Fourth, the presence of ω as a second phase can also result in extra hardening similar to the classical precipitation hardening mechanism.

These combined strengthening mechanisms, with the Hall-Petch mechanism as a major contributor, explain the ultrahigh tensile strength (up to 2130 MPa) observed in the HPT-processed alloy. The presence of nanograins and dislocations can also enhance biocompatibility because cell viability is typically better on defective nanostructured materials, particularly on titanium and Ti-containing alloys [3,22,24,29]. Despite the influence of the HPT treatment on strength and biocompatibility, the formation of nanotubes during anodization appears to be independent of HPT-induced grain refinement. This is reasonable because electrochemical properties are less affected by microstructure, particularly under extreme conditions such as anodization [39-32].

Although microstructural modifications can enhance strength and biocompatibility, they are not effective in controlling the elastic modulus. In this study, phase transformation was used to control the elastic modulus because the elastic modulus is phase-dependent and HPT is a powerful tool to achieve phase transformations in various materials [26,37,45], including titanium [27,43-46] and HEAs [9]. While homogenized Ti-Zr-Hf-Nb-Ta-based HEA exhibits an elastic modulus of 97±4 GPa, which is consistent with the literature [35], the elastic modulus decreases to 69 ±2 GPa after HPT processing, which is appreciably less than the elastic modulus of titanium observed in the current work and reported in the literature [54,55]. The reduction of the elastic modulus should be attributed to the presence of the ω phase, as suggested by first-principles calculations [56]. While the ω phase is generally considered a brittle phase [43-45], its presence in the current HEA does not result in a brittle behavior, and high plasticity values in the range of 13-85% are achieved. Earlier studies suggested that the metastable phases can create mechanical instability and enhance the ductility of HEAs [48]. An earlier study also showed that activation of nanotwins accompanied by BCC-to-ω phase transformation could result in a high elongation in TiZrHfNbTa even at cryogenic temperatures [47]. The formation of nanotwins after $N = 1/8$ turns, which was confirmed in Fig. 3 using TEM, was suggested in different publications as an approach to obtain high strength and high plasticity [37-42]; however, such nanotwins could not be detected in this study after $N = 1$ and 10 turns.

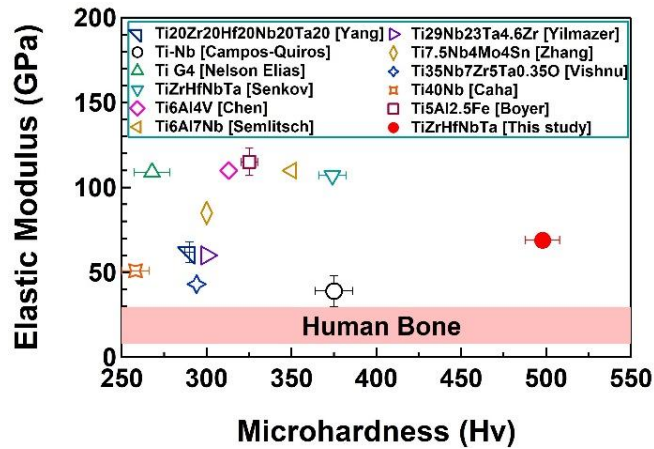

Figure 13. Elastic modulus plotted against hardness for Ti-Zr-Hf-Nb-Ta-based high-entropy alloy in comparison with other biomaterials [57-62] and human bones [63,64].

Taken all together, a combination of high-entropy core effects, lattice defects, nanograins and metastable phase formation contributes to enhanced strength, good plasticity, low elastic modulus and good biocompatibility of HPT-processed HEA. This integration of concepts of HEA and SPD (either as a synthesis method for metals and ceramics like TiZrHfNbTa [71], TiZrHfNbTaO$_{11}$ [72] and TiZrHfNbTaO$_6$N$_3$ [72], or a processing tool [23,73]) is expected to faciliate the discovery of a wider range of biomaterials in the future.

## 5. Conclusions

In the current investigation, high-pressure torsion was employed to process a high-entropy alloy containing non-toxic and allergy-free elements to achieve a good synergy between mechanical properties, anodization and biocompatibility. The Ti-Zr-Hf-Nb-Ta-based alloy, rich in nanograins, dislocations and metastable ω phase, shows superior properties, including large microhardness reaching 500 Hv, large tensile strength up to 2130 MPa, low elastic modulus down to 69 GPa and excellent anodization and biocompatibility compared to pure titanium. The high strength and biocompatibility are due to the nanostructural features of this alloy, while its reduced elastic modulus is due to the metastable phase formation. The results underscore the high potential of severely deformed HEAs for applications as biomaterials.

## CRediT Authorship Contribution Statement

All authors: Conceptualization, Methodology, Investigation, Validation, Writing – review & editing.

## Declaration of competing interest

The authors declare no competing interests that could have influenced the results reported in the current article.

## Acknowledgments

This study is supported partly by Grants-in-Aid from the Japan Society for the Promotion of Science (JP22K18737), and partly by the São Paulo Research Foundation FAPESP (PI-Grant #2022/03024-7 and Scholarship #2023/07884-3).

## Data Availability

The data will be made available upon request from the corresponding author.